\documentclass[submission]{eptcs}
\providecommand{\event}{PrePost 2016}
\usepackage[english]{babel}
\usepackage[utf8x]{inputenc}
\usepackage[T1]{fontenc} 
\usepackage{graphicx}
\usepackage{amsmath}
\usepackage{amssymb}
\usepackage{bm}
\usepackage{color}
\usepackage{hyperref}
\usepackage{subfig}
\usepackage{enumitem}
\usepackage{tikz}
\usepackage{xspace}

\usetikzlibrary{positioning, chains,calc,shapes,automata,arrows,matrix,backgrounds,fit,shadows,patterns}

\usepackage{tubscolors}
\setlist[itemize]{topsep=0.75ex,itemsep=0.25ex,partopsep=0ex,parsep=0.5ex}
\setlist[enumerate]{topsep=0.75ex,itemsep=0.25ex,partopsep=0ex,parsep=0.5ex}

\tikzstyle{shadedBlue} = [top color=tuDarkBlue20, bottom color=tuDarkBlue40, draw=tuDarkBlue80, thick]
\tikzstyle{shadedRed} = [top color=tuRed20, bottom color=tuRed40, draw=tuRed80, thick]
\tikzstyle{shadedOrange}  = [top color=white, bottom color=tuOrange40,  draw=tuOrange100, thick]
\tikzstyle{shadedGreen}  = [top color=white, bottom color=tuGreen40,  draw=tuGreen100, thick]
\tikzstyle{shadedYellow} = [top color=white, bottom color=tuYellow40,  draw=tuYellow100, thick]
\tikzstyle{shadedGray} = [top color=white, bottom color=tuGray20, draw=tuGray60, thick]
\tikzstyle{shadedGrayLight} = [top color=tuBlack!5, bottom color=tuBlack!10, draw=tuBlack!30, thick]

\tikzstyle{shadow} = [drop shadow={opacity=.5,shadow xshift=.3ex,shadow yshift=-.3ex}]
\tikzstyle{triangle} = [isosceles triangle,isosceles triangle stretches]
\tikzstyle{label-it} = [font=\itshape]

\tikzstyle{block} = [draw, shadedBlue, rectangle, rounded corners, minimum height=2em, minimum width=5em,shadow]
\tikzstyle{smallblock} = [draw, shadedBlue, rectangle, rounded corners, minimum height=1em, minimum width=2em,shadow]
\tikzstyle{outerblock} = [draw, shadedGrayLight, draw=tuGray80, rectangle, rounded corners, minimum height=2em, minimum width=5em,shadow]
\tikzstyle{bubble} = [fill=black,shadow,circle,draw=black,inner sep=0pt,minimum size=5pt]
\tikzstyle{memory} = [cylinder, shape border rotate=90, aspect=.4, shadedGrayLight, minimum width=5em, shadow]

\tikzstyle{inheritArrow} = [-open triangle 60,thick]
\tikzstyle{kompArrow}    = [diamond-,thick]

\tikzstyle{flowDecision} = [diamond, draw, shadedRed, text badly centered, inner sep=0pt,shadow]
\tikzstyle{flowBlock} = [rectangle, draw, shadedBlue, text centered, rounded corners, minimum height=2em,shadow]
\tikzstyle{bgBox} = [rectangle, draw, shadedGrayLight, text centered, rounded corners=3mm, shadow, inner sep=10pt]

\tikzstyle{blockarrow} = [draw, thick, single arrow, minimum height=3em]

\newtheorem{definition}{Definition}

\usepackage{listings}

\newcommand{\sys}{\mathit{sys}}

\newcommand{\config}{\mathit{config}}
\newcommand{\soft}{\mathit{software}}
\newcommand{\plat}{\mathit{platform}}

\newcommand{\contract}{\mathcal{C}}

\newcommand{\changeType}{\mathit{changeType}}

\newcommand{\add}{\mathit{add}}
\newcommand{\remove}{\mathit{remove}}
\newcommand{\update}{\mathit{update}}

\newcommand{\eg}{e.\,g.,\ }
\newcommand{\ie}{i.\,e.,\ }

\newcommand{\connection}{\mathit{SC}}
\newcommand{\mapping}{\mathit{M}}
\newcommand{\priorities}{\mathnormal{\Pi}}

\newcommand{\pa}{\texttt{\small park\_assist}}
\newcommand{\sor}{\texttt{\small ob\-ject\_recog\-ni\-tion}}
\newcommand{\om}{\texttt{\small object\_masking}}
\newcommand{\steering}{\texttt{\small steering}}
\newcommand{\tc}{\texttt{\small trajectory\_calculation}}

\newcommand{\ctc}{T}

\newcommand{\orget}{\texttt{\small object\_recognition\_get}}
\newcommand{\tcget}{\texttt{\small tra\-jec\-to\-ry\_cal\-cu\-la\-tion\_get}}
\newcommand{\tcinit}{\texttt{\small tra\-jec\-to\-ry\_cal\-cu\-la\-tion\_init}}

\usepackage[nolist]{acronym}
\begin{acronym}
	\acro{CCC}{Controlling Concurrent Change}
	\acro{RTE}{Runtime Environment}
	\newacroindefinite{RTE}{an}{a}
	\acro{RPC}{Remote Procedure Call}
	\newacroindefinite{RPC}{an}{a}
	\acro{MCC}{Multi-Change Controller}
\end{acronym}

\begin{document}

\title{Using Multi-Viewpoint Contracts\\ for Negotiation of Embedded
  Software Updates\vspace{\baselineskip}}

 \author{Sönke Holthusen$^1$, Sophie Quinton$^2$, Ina Schaefer$^1$,
   Johannes Schlatow$^1$, Martin Wegner$^1$
   \institute{$^1$TU Braunschweig}
   \email{\{s.holthusen, i.schaefer\}@tu-bs.de, schlatow@ida.ing.tu-bs.de, wegner@ibr.cs.tu-bs.de}
   \and
   \vspace{-.8cm}
    \institute{$^2$Inria Grenoble - Rhône-Alpes}
	\email{sophie.quinton@inria.fr}
 }

\def\authorrunning{S. Holthusen, S. Quinton, I. Schaefer, J. Schlatow, M. Wegner}
\def\titlerunning{Using Multi-Viewpoint Contracts for Negotiation of Embedded Software Updates}

\maketitle              

\begin{abstract}
In this paper we address the issue of change after deployment in
safety-critical embedded system applications. Our goal is to
substitute lab-based verification with in-field formal analysis to
determine whether an update may be safely applied. This is challenging
because it requires an automated process able to handle multiple
viewpoints such as functional correctness, timing, etc. For this
purpose, we propose an original methodology for contract-based
negotiation of software updates. The use of contracts allows us to
cleanly split the verification effort between the lab and the
field. In addition, we show how to rely on existing viewpoint-specific
methods for update negotiation. We illustrate our approach on a
concrete example inspired by the automotive domain.
\end{abstract}
\renewcommand{\arraystretch}{1.25}

\section{Introduction}
\label{sec:introduction}
Critical embedded systems currently offer limited support for software
updates. In the automotive domain, this leads to outdated software
being used on mechanically durable vehicles. Adaptable vehicle
electronics could be used, e.\,g., for increasing safety via software
patches, or to limit costs by a better usage of computational
resources. In a different context, aerospace computing platforms are
currently migrating from fixed, ground-verified configurations to
re-configurable platforms, and from functionally separated solutions
to integrated systems with the capability to change during
operations. These two examples illustrate the fact that we need to
support {\em flexible} adaptation of safety-critical embedded systems
via continuous change after deployment.  

At the moment software updates in critical embedded systems are
prepared via intensive verification and tests performed in the
lab. Unfortunately, this lab-based verification process is becoming
increasingly difficult, in particular because of complex platform
dependencies between changing applications---for example through
timing, fault handling, security mechanisms, etc. Every particular
system configuration is different and there are just too many of them
to perform exhaustive a priori verification, even using abstraction
techniques. For this reason, our goal is to use {\em in-field} formal
analysis to determine whether a specific update in a given system
configuration may be safely applied.

Although promising, in-field verification of software updates for
safety-critical embedded systems is also challenging because it
requires an automated process able to handle multiple viewpoints such
as functional correctness, timing or security, with access to limited
resources. For this purpose, we propose a {\em contract}-based
methodology. Such an approach is particularly well-suited to cleanly
split the verification effort between the lab (pre deployment) and the
field (post deployment). In
addition, it lends itself to our original approach whose specificity
is to address multi-viewpoint verification based on a combination of
existing viewpoint-specific methods. Our work is motivated by the
project \emph{Controlling Concurrent Change}
(CCC)\footnote{\url{http://ccc-project.org/}}, which addresses new
methods to develop and control embedded system platforms
integrating changing applications under high requirements to real-time,
safety, availability, and security. The methodology is currently being
implemented as a complete tool chain. We strive to present our current
results
both at a high level of abstraction, so that our results can be reused
or adapted to other contexts, and at a lower level of abstraction so
that our theory matches the practical needs of the CCC project. This
dual approach is reflected in this paper. \medskip

This paper is organized as follows. Section~\ref{sec:contracting}
introduces the general methodology that we
propose. Section~\ref{sec:ccc} then presents the actual context in
which we develop this approach. In Section~\ref{sec:example} we show
the effectiveness of our methodology on a concrete example in the
automotive domain. Finally, Section~\ref{sec:discussion} discusses the
state of the art and Section~\ref{sec:conclusion} concludes.

%

%
\section{Contracting for software update negotiation}
\label{sec:contracting}
We present here our general approach to in-field negotiation of software
updates.

\subsection{Problem statement}
Let us state first that, independent of verification, software updates
require a component-based approach. The overall objective of in-field
negotiation is then to guarantee that conformance to system
requirements is preserved during the update. It is however likely that
establishing conformance monolithically will not be feasible in
practice; hence the use of contracting, which clearly splits
responsibilities between a component and its environment so as to make
design and verification easier. As a result, our methodology assumes
that the following primitives are given.

\paragraph{Component framework.} We suppose a given notion of
component so that one can specify requirements and verify that a given
system conforms to them.
\begin{itemize}
\item {\em component}: In this paper we address software updates so
  the term component refers to software components. How these
  components are specified depends on the actual framework, but at
  this point we do not make any assumption about them. A set of
  components form what we call a {\em software model}.
\item {\em system model}: A software model alone is not sufficient to
  describe a system if one wants to verify non-functional
  properties. Therefore, we define a system model as a triple made of:
  \begin{itemize}
  \item a software model;
  \item a {\em platform model}, which is not subject to change;
  \item a {\em configuration} that describes all runtime parameters relating components and platform which may be changed during negotiation. 
  \end{itemize}
\item {\em system requirements}: The whole verification problem supposes that there exists some formal representation of the expected behavior of a system, which we suppose given as a set of requirements. In addition, we need a formal relation between systems and requirements, that we call {\em conformance}, to allow establishing that a given system conforms to its specified requirements. 
\end{itemize}

\paragraph{Contract theory.}
We further assume primitives relating to contracting.

\begin{itemize}
\item {\em contract}: A contract for a component $K$ is a pair $(A,G)$
  consisting of an {\em assumption} $A$ on the system in which $K$
  will execute and a {\em guarantee} $G$ on the way $K$ will behave in
  such a system.
\item {\em contract satisfaction}: Components and contracts are
  related via a contract satisfaction relation. Intuitively, a
  component {\em satisfies} a contract $\contract=(A,G)$ if, assuming
  that $A$ holds, then $G$ holds, too.
\item {\em compatibility of contracts}: This concept is derived from
  the above primitives. A set of contracts is {\em compatible with
    respect to} a platform model if there exists a {\em feasible
    configuration}, \ie a configuration such that, in any system made
  of components satisfying the given contracts, if assumptions on the
  environment of the system hold then all assumptions and therefore
  all guarantees hold.
\end{itemize}
Contract compatibility is the cornerstone of the contract-based
approach as it does not require the actual components to be available,
only their contracts. Because of that, we use from now on the term
{\em software model} to refer to the set of contracts abstracting the
components.

Besides, note that an additional verification
step is in principle needed after compatibility has been established
in order to derive conformance to system requirements from assumptions
and guarantees. In our case, system requirements are always expressed
directly as assumptions or guarantees so we will largely ignore this
step in the rest of the paper.\medskip

In a context where a component framework and a corresponding contract
theory have been developed, we can now formulate our problem as follows.

\begin{definition} An {\em update request} is a pair $(\changeType,\contract)$ where
\begin{itemize}
\item $\changeType$ describes the change to undertake which is one of
  the following $\{\add,\remove,\update\}$;
\item $\contract$ is the contract attached to the component undergoing
  change.
\end{itemize}
\end{definition}
The objective of the {\em negotiation process} is to determine whether
the update request can be granted and if that is the case then for
which configuration.

\begin{definition}
\label{def:negotiation}
A {\em negotiation process} is
a function taking a system model
$\sys = (\soft, \plat, \config)$ and an update request
$(\changeType,\contract)$ as parameters and returning an {\em answer}:
either {\em no}, or {\em yes} along with a feasible configuration $c$.
\end{definition}
This means that negotiation is a synthesis problem: we need to
synthesize a configuration that makes all contracts compatible.

\subsection{Design and verification flow}
\label{sec:design}
The use of contracts effectively splits the verification process
between the lab and the field, as we show in the following.

\begin{figure}[ht]
	\centering
	\includegraphics[width=.8\textwidth]{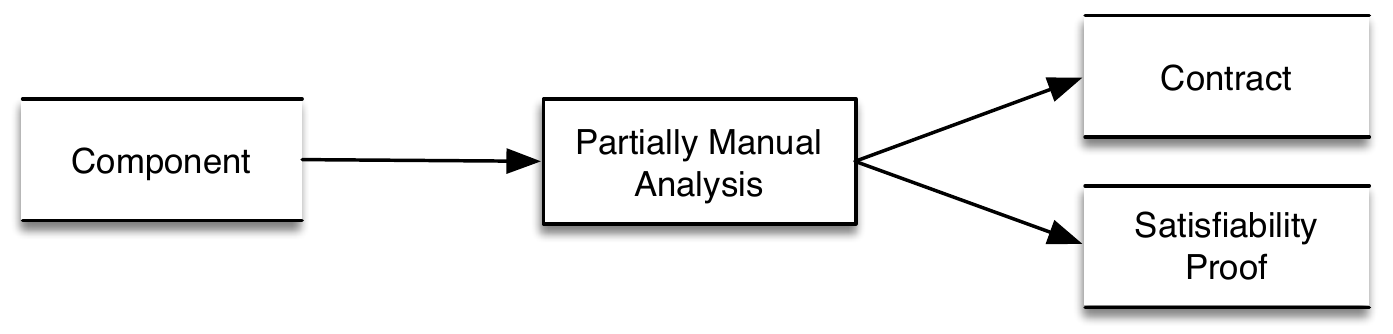}
	\caption{In the lab a contract and a proof for satisfaction is
          generated.}
	\label{fig:labanalysis}
\end{figure}

\paragraph{In the lab.} At design-time (see \autoref{fig:labanalysis}), during software development
and test in the lab, additional data for helping the in-field
negotiation process are prepared and formalized as contracts.  They
will typically include properties of the software components which can
only be obtained using involved verification methods, e.\,g.,
requiring manual intervention such as theorem proving, or intensive
computation, for example in the case of model checking. Assumptions on
the environment in which the system will function may also be
specified. Contract satisfaction is then established in the lab, as
both the component and its contract are available.  This step is
performed for every component independently and saves the effort to
verify every possible combination of components and possibly versions
of components.

\paragraph{In the field.} After deployment, whenever an update is requested, formal
verification is performed to determine whether the update may take
place without risking a violation of system requirements. This step
precedes the actual update and the system architecture is designed to
ensure that it does not disrupt the running system. A
negotiation process takes place, aiming at finding a system
configuration which will guarantee that all system requirements will
be satisfied after the update. This is the challenge of the approach as
this synthesis and verification phase must be performed automatically,
with limited memory and in reasonable time. Updates will typically
be performed while the system is suspended (e.\,g., overnight while the
car is parked) so that the time constraint exists but is less
tight than the memory constraint.

\begin{figure}[ht]
	\centering
	\includegraphics[width=.6\textwidth]{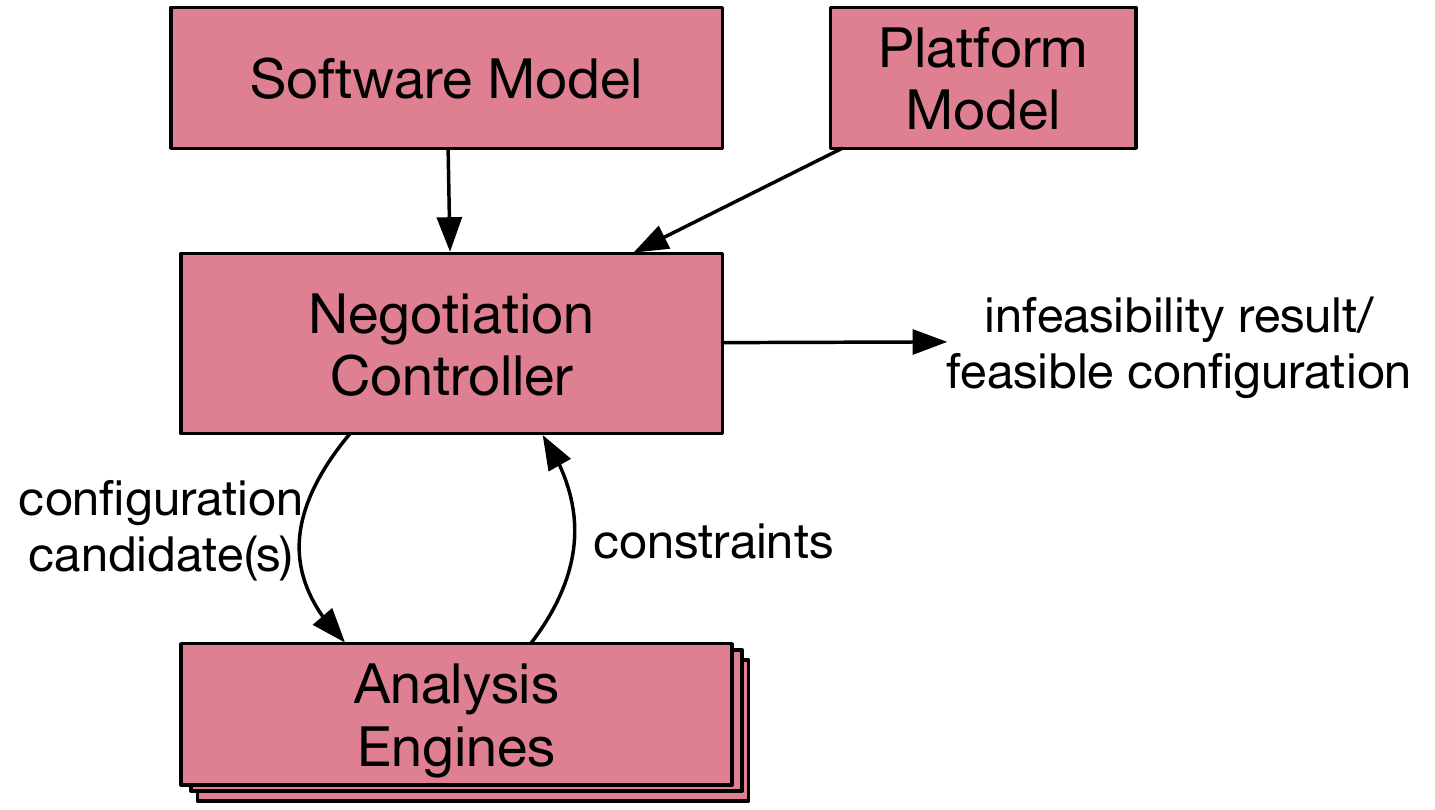}
	\caption{With the contracts from the software model and the
	available hardware resources from the platform model, the
	contract negotiation takes place in field.}
	\label{fig:fieldanalysis}
\end{figure}

To succeed, the component in charge of the negotiation, which we call
the {\em Negotiation Controller} (NC), can rely on the contracts that
have been prepared at design time. A first specificity of our
approach is that we do not try to tackle all viewpoints together as in
\eg\cite{BenvenisteCFMPS07}. Instead, we rely on {\em
  viewpoint-specific analysis engines}, each of which uses a model
that is best adapted to the particular viewpoint it is dealing with.
This also means, we do not encode the search for a configuration
into one big constraint-solving problem including all viewpoints.
We reduce the possible variables for the main problem to a minimum,
while the viewpoints can solve viewpoint-specific problems on their
current partial configuration.
Besides, the brute-force approach that involves checking all possible
configurations until a feasible one is found is not viable due to the
size of the configuration space. To tackle this, a second specificity
of our approach is that we rely on added capabilities of the
viewpoint-specific analysis engines. Indeed, our experience is that
these engines, although used in general for checking one given
configuration, can be extended to provide information about sets of
configurations. Our strategy is therefore to start from an
over-approximation of the set of feasible configurations and
iteratively update this set based on the results provided by the
analysis engines. More precisely we proceed as shown in \autoref{fig:fieldanalysis}.

  \begin{enumerate}
  \item The NC computes a set of {\em configuration candidates}
    which is represented as a set of constraints.
  \item The NC then picks one candidate and triggers the
    viewpoint-specific engines so that they check if that
    candidate configuration is feasible w.\,r.\,t. the considered viewpoint
    --- and as such a witness to the
    software model being compatible with respect to the platform
    model.
  \item If the configuration is not feasible, one engine returns a set
    of additional constraints to the NC. If the configuration is
    feasible for all viewpoints then the update can be accepted.
  \item The last two steps repeat until the NC can no longer find a
    candidate or if a feasible configuration has been found.
\end{enumerate}
In the rest of this paper, we show in more detail how we follow this
approach in a specific, applied context.

\section{Software update negotiation in the CCC project}
\label{sec:ccc}
In this section, we present the architecture that we are currently
developing to support negotiation in the context of the CCC
project. Our goal in this section is to define all the concepts used
in Definition~\ref{def:negotiation}, namely: a software model
represented as a set of contracts, a platform model and a notion of
configuration --- and to show how we use them to perform contract
negotiation.\vspace{-0.5\baselineskip}

\paragraph{Platform architecture.}
Similar to integrated architectures like AUTOSAR~\cite{AUTOSAR}
or ARINC 653, our approach offers a clear separation
between software/application and hardware/platform components through
a \ac{RTE}. We use the Genode OS Framework~\cite{GENODE} as \ac{RTE}, which
allows for the design of component-based systems where access control is
applied via so-called capabilities.
\label{sec:architecture}
Our architecture complements the \ac{RTE} with a so-called {\em \ac{MCC}}
which acts as negotiation controller. Similar to the EPOC~\cite{NeukirchnerSE11}
project, the \ac{MCC} is divided into a {\em model domain}, in which
analyses are performed without actually modifying the system, and an
{\em execution domain}, in which the update is implemented. In this
paper, we focus on the model
domain.\vspace{-0.5\baselineskip} 

\paragraph{Component framework.}
\label{sec:components}
Our software is built as a set of {\em functions} made of {\em
  components}. For convenience, we only consider here functions with a single component
so most of the time, we will abstract functions away. Components are
interconnected via {\em services}: components can offer services to
each other, or require services from one another. Matching between
offered and required services is made via comparison of the
corresponding {\em service interfaces}, which describe the services'
expected type of communication.

Communication between components can
be performed synchronously via {\em \acp{RPC}} or asynchronously using
{\em signals}. More precisely, \acp{RPC} allow passing arguments to
the component being called, but the caller cannot continue its
execution until the callee has finished handling the request. In
contrast, signals can be used asynchronously, only notifying the
callee and ending the communication afterwards.

Components are implemented as one or more \emph{threads} such that
each thread can be activated by: (1) another component requiring a
service via an RPC or a signal; (2) another thread of the same component; or (3) a
timer.\vspace{-0.5\baselineskip}

\paragraph{Requirements.} Although the scope of the CCC
project is larger, we consider in this paper the following viewpoints.
\begin{itemize}
\item {\em functional dependencies}: This viewpoint addresses issues
  related to the structural information presented above, e.\,g., to guarantee that all required
  services are indeed provided.
\item {\em timing}: The timing viewpoint is typically concerned with
  establishing that a function reacts to an input within a specified delay.
\item {\em functional correctness}: Functional correctness focuses on
  control flow properties, e.\,g., expressing that a given service
  cannot be called before some initialization step has been performed.
\end{itemize}
All the above system requirements can be expressed at the component or
function level so we specify them directly in our
contracts. Conformance to these requirements is thus established if
contract compatibility is guaranteed.\vspace{-0.5\baselineskip}

\paragraph*{Contract framework.}
To express assumptions and guarantees related to our three viewpoints
we have developed a contract language capturing all the relevant
information. Rather than using a formal definition, in the following we choose to
describe our language through an example.\vspace{\baselineskip}

\noindent\emph{Functional dependencies.} \autoref{listing:tc} shows the contract for a component \ctc, which
requires a service \sor\ (line 3) and provides a service \tc\ (line
4). The names of the required and provided services are IDs, linking
to a more detailed service description in a global repository. Knowing
the required and provided services of all components, functional
dependencies can be resolved~\cite{SchlatowME15}.\vspace{\baselineskip}

\noindent\emph{Timing.} The timing requirement is specified at lines 20--22
and states that the RPC call to \sor\ must complete within 100 time
units. In addition, to enable timing analysis~\cite{HeniaHJ+05},
contracts must contain enough information to extract a task
graph~\cite{HeniaRSGP15}. A task graph is a directed graph in which
nodes represent tasks and edges represent the asynchronous (or
synchronous) activations/calls and returns. For that purpose the
thread structure as well the nature of communication (\ac{RPC} or
signal) must be known. In our example, the component is made of two
threads called \tcget\ and \tcinit. The former is triggered via
\ac{RPC} (line 7 in the listing). When activated it first proceeds
with executing some local processing task which takes between 1 and 5
time units to complete on a computing resource of type \texttt{\small
  CPU\_type\_1} (lines 8 to 10). It then performs an RPC call to use a
service (line 11) and finally executes another task before completing
(lines 12--13).\medskip

\begin{lstlisting}[caption=An example contract.
%\footnote{To reduce the size of  the listing and because the indentation ensures syntactical  integrity, we removed the parentheses. We also reduced some  information not necessary for the explanation.}
, captionpos=b, label=listing:tc]
component T
	services 
		requires object_recognition
		provides trajectory_calculation
	threads 
		thread trajectory_calculation_get 
			on RPC trajectory_calculation.get() 
				task tc1 
					onto CPU_type_1 
						wcet=5 bcet=1
				RPC object_recognition.get()
				task tc2
					onto CPU_type_1 
						wcet=5 bcet=1
		thread trajectory_calculation_init 
			on RPC trajectory_calculation.init() 
				task tci 
					onto CPU_type_1 
						wcet=10 bcet=5
	timings
		timing 100
			object_recognition.get()
control_flow 
  not trajectory_calculation.get()
    until trajectory_calculation.init() 
\end{lstlisting}

\noindent\emph{Functional correctness.} Finally, the functional
requirement is specified at lines 23--25. It requests that the trajectory
calculation must always be initialized before it is used. Such control
flow properties can be easily checked on a control flow graph
extracted from the contracts.\medskip

Note that our contracts do not explicitly distinguish assumptions from
guarantees. In addition, a significant part of the guarantees hold
independent of the assumptions and can thus be regarded as part of a
specification rather than a contract. We have chosen this presentation
to improve readability. In the contract for component \ctc, the
assumptions correspond to the three viewpoint-specific
requirements. The most notable guarantees regard worst-case execution
time bounds, which can only be formally established using involved
verification techniques including abstract
interpretation~\cite{WilhelmEEHTWBFHMMPPSS08}.

\paragraph{System model.} Let us synthesize the information
presented above to define our system model. Remember that a system
model is made of a software model, a platform model and a
configuration.
\begin{itemize}
\item {\em software model}: this is the set of contracts used as an
  abstraction layer on top of
  components. 
  Note that only the components that are actually part of the chosen
  configuration will be loaded when a new configuration is set up.
\item {\em platform model}: for contract negotiation we need some
  abstraction of the platform as well. The only non-functional
  viewpoint that we address here is timing, so information about the
  available processing resources and their scheduling policy is
  sufficient, as will be detailed later. As already mentioned, we
  assume that the platform model is fixed.
\item {\em configuration}: this describes how the selected software
  components are connected to each other and to the hardware. It
  consists of a set of connections (functional dependencies) between
  offered and required services, a mapping of tasks to the platform
  resources, and priorities assigned to tasks.
\end{itemize}

Denote $\soft$ the set of software contracts that make up the software
model. The set of services (and whether they are required or offered
by a given component) as well as the set of tasks that correspond to a
given component can easily be extracted from the component's contract.

Denote $\plat$ the set of processing resources that constitute the
platform model. In our case all resources have the same scheduling
policy, namely static-priority preemptive, so the only additional
information we need about them is their type in order to guarantee an
appropriate mapping of tasks.

\begin{definition}A {\em configuration} $\config$ is a tuple
  $(C, \connection, \mapping, \priorities)$ where:
\begin{itemize}
\item $C\subseteq \soft$ represents the set of selected components,
  \ie components which must be loaded and run on the platform.
\item $\connection$ is a set of {\em service connections}, \ie of
  triples $(c_1, s, c_2)$ with $c_1, c_2 \in C$ representing a connection between two
  components $c_1$ and $c_2$ via a service $s$ .
\item $\mapping$ is a function associating to each task the resource
  onto which it is mapped.
\item $\priorities$ is a total priority order on tasks.
\end{itemize}
\end{definition}
Note that not all these configurations are {\em well-formed} as we
discuss below.

\paragraph{Negotiation process.}
In this section we formally define how the configuration space is
constrained during the contract negotiation. The objective of the
negotiation process is to find a feasible configuration. To achieve
this the \ac{MCC} picks a configuration from a set of candidates
called the {\em configuration space} and runs it through the
viewpoint-specific analysis engines to determine whether it is
feasible. Initially the configuration space contains all {\em
  well-formed} configurations (see below). It is then iteratively
restricted based on the feedback provided by the viewpoint-specific
analysis engines.

It has been shown in~\cite{SchlatowME15} that the
functional dependency resolution problem can be formulated entirely as
a set of constraints, so there is no need for a specific, dedicated analysis
engine. Such constraints include in particular:
\begin{enumerate}
\item For all $(c_1, s, c_2)\in \connection$, $c_1\neq c_2$ and the
  service $s$ is indeed required by component $c_1$ and offered by
  component $c_2$.
\item For every selected component $c_1\in C$ requiring a service $s$,
  there is a unique selected component $c_2\in C$ such that
  $(c_1, s, c_2)\in \connection$.
\end{enumerate}
Regarding mapping and priority assignment, not all configurations
are well-formed either as the following constraints must be met:
\begin{enumerate}
\item[3.] A task $\tau$ can only be mapped onto a resource $r$ if $r$
  is of the type for which execution time bounds are provided in the
  contract corresponding to $\tau$.
\item[4.] Tasks from the same thread must have the same priority.
\end{enumerate}

\begin{definition} A configuration $(C,\connection, \mapping,
  \priorities)$ is {\em well-formed} if and only if it satisfies the
  four above mentioned conditions.
\end{definition}

In contrast, functional correctness may require intricate analysis
which is best kept separate from the main set of constraints. Whenever
a configuration is rejected by the functional correctness analysis
engine checking control flow properties, a new constraint is added to
restrict $\connection$. Such constraint will typically forbid any
configuration in which $c_1$ is selected and $(c_1, s, c_2)$ is a
service connection if it has been established that $c_1$ is
``misusing'' the service offered by $c_2$. Note that this removes a
fairly large set of configuration candidates, namely all possible
mappings and priority assignments for each $\connection$ eliminated.

Finally, timing analysis is the most challenging part of our
negotiation scheme. Encoding the response-time analysis as an ILP (integer linear programming)
problem has been done~\cite{WiederB13}, but at the cost of either a
prohibitive complexity or pessimism, meaning that configurations which
may be feasible are discarded. Our strategy is to perform timing
analysis using state of the art timing analysis tools such as \eg
pyCPA~\cite{pycpa} but to better exploit the provided result. Suppose
for example that timing analysis cannot guarantee that a given task
$\tau$ will meet its deadline in the worst case. This means that for
the current mapping any configuration that keeps the same priority for
$\tau$ and higher-priority tasks will violate $\tau$'s timing
requirement. A full theory of how such constraints can be generated is
in progress and out of the scope of this paper but it is particularly
interesting to remark that such issues have been relatively ignored so
far in the real-time systems research community.

\section{Application to an automotive example}
\label{sec:example}%
Let us now illustrate our approach to contracting for update
negotiation on an example taken from the automotive domain. We
consider a car in which a function for parking assistance is deployed
and show how it can be updated to add a function for lane detection.
We first describe the software components we use and their relevant
properties before elaborating on the negotiation process.

\subsection{System model and viewpoints before the update}
The component P for parking assistance depends on \tc, a service for
calculating a trajectory for the car (i.\,e., a list of vectors to steer
the vehicle). That service, which is provided by component T, itself needs
an \sor\ service to determine which objects must be
avoided. Suppose that two components offering this service are
available (O$_1$ and O$_2$), with O$_2$ additionally offering a
service \om\ for masking objects.

\begin{figure}[ht]
	\centering
        \begin{tikzpicture}[auto, node distance=1.5cm,>=latex',font=\small]

\input{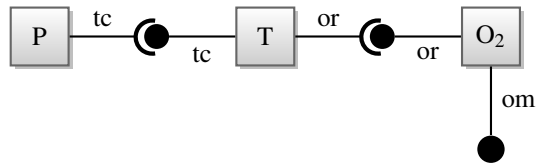}

\node[component](p){P};
\node[service, right of=p](tc){};
\node[component, right of=tc](t){T};
\node[service, right of=t](or){};
\node[component, right of=or](o2){O$_2$};
\node[openservice, below of=o2](om){};

\draw[requires] (p) -- node {tc} (tc);
\draw[provides] (t) -- node {tc} (tc);

\draw[requires] (t) -- node {or} (or);
\draw[provides] (o2) -- node {or} (or);

\draw[provides] (o2) -- node {om} (om);

\end{tikzpicture}
	\caption{Resolution of functional dependencies chosen during
          deployment. Boxes represent components, circles and
          semicircles represent services that are offered and
          required, respectively. Service names are abbreviated to
          their acronym.}
	\label{fig:deployment}
\end{figure}

\paragraph{Initial configuration.}
\label{sec:dependencies}
Assume that during deployment all the above components have been
uploaded onto a unique hardware resource called CPU1. 
\autoref{fig:deployment} shows how functional dependencies are solved
in the current configuration. Note that only P, T and O$_2$ are selected
--- meaning that the code and contract of O$_1$ are available but not
currently in use in the system. The rest of the configuration, namely
task mapping and priority assignment will be discussed later.

\paragraph{Initial requirements.} Let us describe our system
requirements. \autoref{listing:p} shows the contract for component
P. Note that two keywords have not been introduced before, namely {\bf
  \texttt{initialization}} and {\bf \texttt{time}}. The former refers
to an initial {\em operational mode} that precedes the normal mode in
which all other calls are performed. The latter term is used to
introduce an {\em activation pattern} which describes how often a
given thread is activated based on a timer. In our case, thread
\texttt{park\_assist} is activated periodically every 200 time units
with a possible jitter of at most 5 time units.

The contract for T has already been introduced
in~\autoref{listing:tc}. Due to space constraints the contracts for
components O$_1$ and O$_2$ are omitted. In addition to functional
dependency constraints, there are only two timing requirements and one
control flow property to verify:
\begin{enumerate}
\item Each activation of the \pa\ thread must complete within 150 time units ---
  this requirement is specified in~\autoref{listing:p}.
\item A call to \orget\ must return within 100 time units.
\item \tcget\ must not be called before at least one call to \tcinit\ has been made.
\end{enumerate}
The last two requirements are from the contract for component T.

\begin{lstlisting}[float, caption=The contract of P, captionpos=b, label=listing:p]
component P
  services 
    requires trajectory_calculation
  threads
    thread init
      on initialization
        RPC trajectory_calculation.init()
    thread park_assist 
      on time (period=200 jitter=5) 
        task p1 
          onto CPU_type_1
            wcet=3 bcet=1
        RPC trajectory_calculation.get()
        task p2 
          onto CPU_type_1
            wcet=7 bcet=1
  timings 
    timing 150
      park_assist
\end{lstlisting}

\begin{figure}[ht]
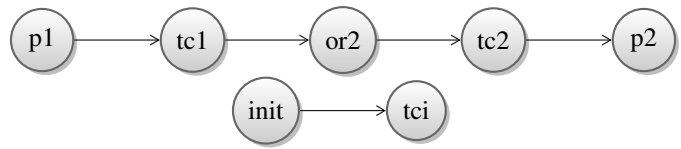

	\centering
	\begin{tikzpicture}[auto, node distance=2cm,>=latex',font=\small]
		\input{tikz/cpastyles}
		\node[task] (p1) {p1};
		\node[task, right of=p1] (tc1) {tc1};
		\node[task, right of=tc1] (o2or) {or2};
		\node[task, right of=o2or] (tc2) {tc2};
		\node[task, right of=tc2] (p2) {p2};

		\draw[activates] (p1) -- (tc1);
		\draw[activates] (tc1) -- 
                (o2or);
		\draw[activates] (o2or) -- (tc2);
		\draw[activates] (tc2) -- (p2);
\end{tikzpicture}\\
  \centering \begin{tikzpicture}[auto, node distance=2cm,>=latex',font=\small]
		\input{tikz/cpastyles}
		\node[task] (rpc1p) {init};
		\node[task, right of=rpc1p] (tci) {tci};

		\draw[activates] (rpc1p) -- (tci);
\end{tikzpicture}
  \caption{The task graph before the update. Note that or2 corresponds
    to the task performing the \sor\ service in component O$_2$.}
	\label{fig:task_graph_before_update}
\end{figure}

\paragraph{Timing viewpoint.}
The task graph presented in~\autoref{fig:task_graph_before_update}
shows two task chains corresponding to the normal mode of operation of
P and its initialization mode, respectively.
This graph has been
obtained by unfolding the task graphs inside the thread description 
with the concrete task graphs of called threads.
This means the \ac{RPC} to \emph{trajectory\_calculation.init()} in \autoref{listing:p}
is replaced by the corresponding task graph from the contract for component
\emph{T} (\autoref{listing:tc}).
Furthermore, the call of \emph{object\_recognition.get()} is replaced by the corresponding task.
This continues until all \acp{RPC} are replaced by tasks.
Note that O$_2$ has a task which is never activated (namely the task
corresponding to the \om\ service) and is thus omitted in the task graph.

Because our platform model consists of only one resource, all tasks
are mapped to CPU1. In presence of several operational modes timing
analysis is performed separately for each mode~\cite{ShaRLR89}. Therefore in the
normal mode of operation only the first chain of tasks is
considered. Our two timing requirements translate at the task level
into latency constraints: one from a call to task p1 to the end of p2;
the other from a call to tc1 to the end of tc2. We will detail how
such requirements are verified in the description of the negotiation
process.\vspace{-0.5\baselineskip}

\paragraph{Control flow graph.} In our example we have chosen a
very simple functional correctness requirement, which can be directly
derived from the semantics of the keyword {\bf
  \texttt{initialization}} in our contract language. We therefore omit
the description of the control flow graph that can be derived from the
contracts and focus on the more intricate timing aspects of our
example. \vspace{-0.5\baselineskip}

\subsection{Update scenario}
Our objective is to add to the system a component L implementing lane
detection, for which it requires services for \sor, \om\ as well as
a service for \steering\ provided by a component S which is also
to be uploaded. We omit the contract for S but show the contract for L
in~\autoref{listing:l}: component L is made of a single thread which
sequentially calls services \sor, \om\ and \steering\ while performing
some internal computation in between calls.\vspace{-0.5\baselineskip}

\begin{lstlisting}[float=ht, caption=The contract of L, captionpos=b, label=listing:l]
component L
  services 
    requires object_masking 
    requires object_recognition
    requires steering
  threads 
    thread lane_assist
      on time (period=100 jitter=5)
        task la1
          onto CPU_type_1
            wcet=3 bcet=1
        RPC object_recognition.get()
        task la2
          onto CPU_type_1
            wcet=3 bcet=1
        RPC object_masking.get()
        task la3
          onto CPU_type_1
            wcet=10 bcet=5
        RPC steering.setAngle(int value)
        task la4
          onto CPU_type_1
            wcet=4 bcet=1
  timings
    timing 75 
      lane_assist
\end{lstlisting}

\subsection{Contract negotiation}
Now we can show how our negotiation process is performed to allow the
addition of the lane detection function.

\begin{figure}[htb]
  \centering \begin{tikzpicture}[auto, node distance=1.5cm,>=latex',font=\small]

\input{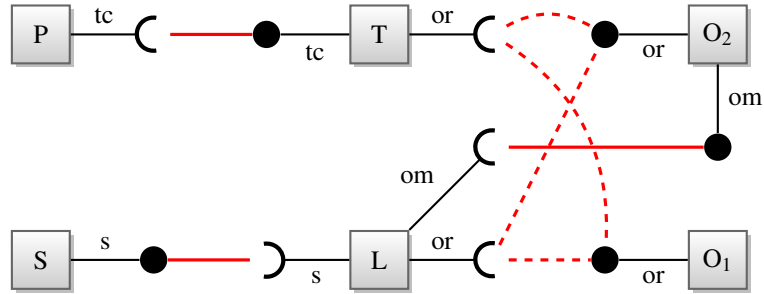}

\node[component](p){P};
\node[openrequirement, right of=p](tcr){};
\draw[requires] (p) -- node {tc} (tcr);

\node[openservice, right of=tcr](tcs){};
\node[component, right of=tcs](t){T};
\draw[provides] (t) -- node {tc} (tcs);

\node[openrequirement, right of=t](orr2){};
\draw[requires] (t) -- node {or} (orr2);


\node[openservice, right of=orr2](ors2){};
\node[component, right of=ors2](o2){O$_2$};
\draw[provides] (o2) -- node {or} (ors2);

\node[openservice, below of=o2](oms){};
\draw[provides] (o2) -- node {om} (oms);

\node[component, below of=oms](o1){O$_1$};
\node[openservice, left of=o1](ors1){};
\draw[provides] (o1) -- node {or} (ors1);

\node[component, below of=t, node distance=3cm](l){L};
\node[openrequirement, right of=l](orr1){};
\node[openrequirement, above of=orr1](omr1){};
\draw[requires] (l) -- node {or} (orr1);
\draw[requires] (l) -- node {om} (omr1);
\node[openrequirement, left of=l, rotate=180](sr){};

\node[component, below of=p, node distance=3cm](s){S};
\node[openservice, right of=s](ss){};
\draw[provides] (s) -- node {s} (ss);
\draw[requires] (l) -- node {s} (sr);

\draw[must] (sr) -- (ss);

\draw[must] (tcr) -- (tcs);
\draw[maybe] (orr2) edge [bend left] (ors2);
\draw[maybe] (orr2) edge [bend left] (ors1);
\draw[must] (omr1) -- (oms);
\draw[maybe] (orr1) -- (ors1);
\draw[maybe] (orr1) -- (ors2);

\end{tikzpicture}
  \caption{Possible partial configurations for our example. Full lines
    denote \emph{must} connections while dashed lines denote
    \emph{may} connections.}
  \label{fig:must}
\end{figure}

\paragraph{Functional Dependency.}
The first step of contract negotiation aims at finding the set of
partial configurations guaranteeing that every service required by a
component in the software model is provided by another component in
that model. In our case we need to account for the services required
by the component to be added to the software model, namely L (note
that S does not require any service). Functional dependency is thus
solved on the software model consisting of components P, T, O$_1$,
O$_2$, S and L. \autoref{fig:must} illustrates the result, as we
explain now. Full lines
between offered and required services represent links for which there
is no alternative. For example, T is the only component offering service
\tc\ required by P, making the connection between the two
mandatory. In contrast, the \sor\ service required by T can be
provided by O$_1$ and O$_2$, which we denote using dashed lines. Any
configuration which matches one of those represented in
\autoref{fig:must} will guarantee compatibility at the functional
dependency viewpoint. For the sake of simplicity we will assume that
the service interfaces for \sor\ prevent O$_1$ and O$_2$ from
offering that service to more than one component. In that case there
are only two possible solutions left.

\paragraph{Functional correctness.}
The part of the control flow graph which is relevant for the
requirement imposed by T does not change after the update so the
functional correctness viewpoint does not need to constrain the
configuration space: any configuration which satisfies the functional
dependency constraints also passes the functional correctness test.

\begin{figure}[ht]
  \centering
  \begin{tikzpicture}[auto, node distance=1.5cm,>=latex',font=\small]
		\input{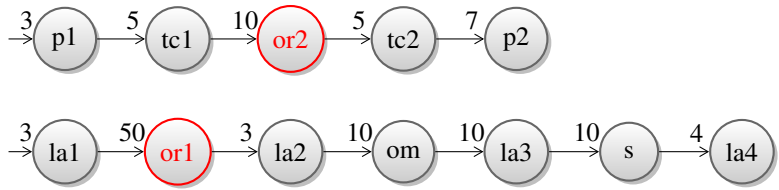}
                \coordinate (p) at (-0.75cm,0);
                \coordinate (l) at (-0.75cm,-1.5cm);

		\node[task] (p1) {p1};
		\node[task, right of=p1] (tc1) {tc1};
		\node[task, red, right of=tc1] (o2or) {or2};
		\node[task, right of=o2or] (tc2) {tc2};
		\node[task, right of=tc2] (p2) {p2};

		\draw[activates] (p) -- (p1) node[near end] {3};

		\draw[activates] (p1) -- (tc1) node[near end] {5};
		\draw[activates] (tc1) -- (o2or) node[near end] {10};
		\draw[activates] (o2or) -- (tc2) node[near end] {5};
		\draw[activates] (tc2) -- (p2) node[near end] {7};

		\node[task, below of=p1, node distance=1.5cm] (la1) {la1};
		\node[task, red, right of=la1] (o1or) {or1};
		\node[task, right of=o1or] (la2) {la2};
		\node[task, right of=la2] (o2om) {om};
		\node[task, right of=o2om] (la3) {la3};
		\node[task, right of=la3](s) {s};
		\node[task, right of=s](la4) {la4};

		\draw[activates] (l) -- (la1) node[near end] {3};

		\draw [activates] (la1) -- (o1or) node[near end] {50};
		\draw [activates] (o1or) -- (la2) node[near end] {3};
		\draw [activates] (la2) -- (o2om) node[near end] {10};
		\draw [activates] (o2om) -- (la3) node[near end] {10};
		\draw [activates] (la3) -- (s) node[near end] {10};
		\draw [activates] (s) -- (la4) node[near end] {4};
\end{tikzpicture}
  \caption{Task graph for one of the two possible functional configurations.}
  \label{fig:tasks}
\end{figure}

\paragraph{Timing.}
The last, more elaborate viewpoint we consider for our example is
timing.
\autoref{fig:tasks} shows the task graph corresponding to one of the
possible functional configurations. The second possible task graph is
obtained by swapping tasks or1 and or2. The values indicated next to
the tasks are their worst-case execution time and therefore a property
of the task and not of the task graph.\vspace{\baselineskip} 

A simple load analysis, performed, \eg using pyCPA\footnote{\url{https://pycpa.readthedocs.org}}, will reject the
functional configuration shown in Figure~\ref{fig:tasks}. The reason
is that the task chain corresponding to the lane assist function is
activated every 100 time units. If or1 is part of this chain, then
that chain requires 90\% of the available CPU time. However, the park
assist chain has a 15\% load (\ie 30 time units every 200 time
units). This means that no matter what the priority assignment is, the
processor will be overloaded. The only option consists in swapping or1
and or2 in the functional configuration.

Our objective is now to find a priority assignment that satisfies the
specified latency constraints. Remember that tasks inherit their
priority from the threads to which they belong. In addition, we do not
assign the same priority to multiple threads as this typically only
adds more pessimism to the analysis result.

Let us focus on the end-to-end latency requirements imposed by P and
L, which are respectively 150 and 75 time units. The constraint on P
is relatively loose and in fact it will be satisfied even if the park
assist function is blocked by the complete lane assist task
chain. Therefore, giving higher priority to the tasks involved in the
lane assist function will always provide a feasible
configuration. Interestingly, it is also possible to establish that
the lane assist chain cannot afford to be blocked by the or1 task. In
that case, a constraint stating that or1 must have lower priority than
all the tasks involved in the lane assist function can be added to
restrict the configuration space.

\subsection{Discussion}
The main purpose of our example was to illustrate how the contract
negotiation process works, and in particular how incompatibility in
one viewpoint can result in additional constraints to the
configuration space. Therefore we kept the example quite simple. Moreover, note
that the methodology is not yet fully implemented: the functional
dependency analysis exists but it is not connected to the timing
analysis engine at the moment. We currently mainly focus on providing useful
timing analysis feedback. 

\section{Related work}
\label{sec:discussion}

Design of complex systems of components using contracts, or
interfaces, has been widely studied since it was introduced by de
Alfaro and Henzinger in~\cite{AlfaroH01}. A great variety of interface
theories have been developed, which mostly focus on incremental
design.

There has been active research lately on contract theories dealing
with multiple viewpoints. For example, Reineke et al. focus on
issues such as {\em consistency} between viewpoints~\cite{ReinekeT14}: if two viewpoints
have some degree of overlap, how is it possible to guarantee that they
do not contradict each other? Perssone et al. give a high-level
classification of model-based approaches to multi-view systems, which
also discusses additional challenges of multi-view modeling~\cite{PerssonTQWBTVD13}:
traceability of information between views, reuse, automation, change
propagation and extendability. In~\cite{PanunzioV14} Panunzio et al. give a more applied
contribution as they propose a component model for separation of
concerns backed by case studies originating from the European Space
Agency. However, none of these results address dynamicity.

Another line of work related to this paper regards
languages for contracts. The BCL language~\cite{FerrantePFMS14} has
now been used in many safety-critical industrial applications. An
alternative is the contract language from~\cite{CimattiT15} which is based on temporal
logic. Note however that our interest is less in the language
aspects than in the methodology.

Changing constrained systems are in the focus of the FRESCOR
project~\cite{SojkaH09}, in which contracts are used to ensure Quality
of Service on the network-layer. However, the project focuses on
bandwidth and latency constraints usually found in multimedia
applications. In this context contracts may be broken (at times),
enabling the test of novel configurations during runtime.  This is not
acceptable when updates have an impact also on safety-critical
functions such as driver assistance systems in a car. The EPOC
project~\cite{NeukirchnerSE11} proposed for the first time a contract-based
admission control framework for safety-critical systems, but it was
restricted to timing aspects and could not handle multiple viewpoints.

\section{Conclusion}
\label{sec:conclusion}

It has become increasingly difficult to test and verify software
updates for embedded systems, in particular because of the complex
platform dependencies which exist between software components. In this
paper we have presented a methodology to split the necessary effort
between lab and field.

Performing in-field formal analysis to determine whether an update may
be safely applied is challenging because it requires an automated
process able to handle multiple viewpoints such as functional
correctness, timing, etc. We have proposed an original methodology
based on contract negotiation which enables this. In particular,
instead of addressing all viewpoints together we rely on
viewpoint-specific analysis engines, each of which uses a model that
is best adapted to the particular viewpoint it is dealing with. A
second specificity of our approach is that we rely on added
capabilities of these viewpoint-specific analysis engines.

Finally, we have illustrated the benefits of our approach on a
realistic scenario from the automotive domain, which we unfold from
the specification of contracts until a safe update has been found.

\bibliographystyle{splncs03}
\bibliography{bib}

\end{document}